\documentclass[aps,prd,twocolumn,nofootinbib,longbibliography,10pt]{revtex4-2}
\usepackage{etoolbox}
\usepackage{dcolumn}
\usepackage{amsmath,amssymb,amsfonts,mathtools}
\usepackage{mathrsfs,bbold}
\usepackage{graphicx}
\usepackage[colorlinks=true,urlcolor=blue,citecolor=red,linkcolor=blue]{hyperref}
\usepackage{accents}
\newlength{\dhatheight}
\usepackage{natbib}
\usepackage{wrapfig}
\usepackage{flushend,BOONDOX-cal,BOONDOX-frak}

\makeatletter
\newsavebox{\@brx}
\newcommand{\llangle}[1][]{\savebox{\@brx}{\(\m@th{#1\langle}\)}%
  \mathopen{\copy\@brx\kern-0.5\wd\@brx\usebox{\@brx}}}
\newcommand{\rrangle}[1][]{\savebox{\@brx}{\(\m@th{#1\rangle}\)}%
  \mathclose{\copy\@brx\kern-0.5\wd\@brx\usebox{\@brx}}}
\makeatother
\begin{document}
\title{\textbf{Atom falling into a quantum corrected charged black hole and HBAR entropy}}
\author{Arpita Jana}
\email{janaarpita2001@gmail.com}
\affiliation{Department of Astrophysics and High Energy Physics, S. N. Bose National Centre for Basic Sciences, JD Block, Sector-III, Salt Lake City, Kolkata-700 106, India}
\author{Soham Sen}
\email{sensohomhary@gmail.com}
\affiliation{Department of Astrophysics and High Energy Physics, S. N. Bose National Centre for Basic Sciences, JD Block, Sector-III, Salt Lake City, Kolkata-700 106, India}
\author{Sunandan Gangopadhyay}
\email{sunandan.gangopadhyay@gmail.com}
\affiliation{Department of Astrophysics and High Energy Physics, S. N. Bose National Centre for Basic Sciences, JD Block, Sector-III, Salt Lake City, Kolkata-700 106, India}
\begin{abstract}
\noindent   In an earlier analysis \href{https://link.aps.org/doi/10.1103/PhysRevD.105.085007}{Phys. Rev. D 105 (2022) 085007}, we have explored the event of acceleration radiation for an atom freely falling into the event horizon of a quantum-corrected Schwarzschild black hole. We want to explore the acceleration-radiation when the atom is freely falling into the event horizon of a charged quantum-corrected black hole. We consider the quantum effects of the electromagnetic field along with the gravitational field in an asymptotic safety regime. Introducing the quantum improved Reisner-Nordstr\"{o}m metric, we have calculated the excitation probability of a two-level atom freely falling into the event horizon of quantum improved charged black hole. Recently, in the case of the braneworld black hole (where the tidal charge has the same dimension as the square of the charge of a Reissner-Nordstr\"{o}m black hole in natural units), we have observed from the form of the transition probability that the temperature will have no contribution in the first order of the tidal charge. We observe that for a quantum corrected Reissner-Nordstr\"{o}m black hole, there is a second-order contribution in the charge parameter in the temperature that can be read off from the transition probability. Next, we calculate the HBAR entropy in this thought experiment and show that this entropy has a leading order Bekenstein-Hawking entropy term along with some higher order correction terms involving logarithmic as well as fractional terms of the black hole area due to infalling photons.  We have finally investigated the validity of Wien's displacement law and compared the critical value of the field wavelength with the general Schwarzschild black hole and its corresponding quantum-corrected case.
\end{abstract}
\maketitle
\section{Introduction}

\noindent The theory of general relativity, constructed and developed by Albert Einstein in the first quarter of the twentieth century \cite{Einstein1,Einstein2}, is considered to be one of the most accurate classical theories explaining the laws of nature. Later on with the development of quantum mechanics, people tried to probe the general relativistic arguments in very small length scales. The  theory of general relativity breaks down at very small length-scales due to its non-renormalizability \cite{DesserNieuwenhuizen,Klauder} and when quantized it leads to several discrepancies because of the inherent non-linearity of the theory. This problem pushed researchers for almost a century to search for a correct and renormalizable quantum theory of gravity. Quantum mechanical effects in the general theory of relativity can be incorporated in several ways. One of the ways to do this is the asymptotic safety program which is based on the renormalization group approach initially introduced in \cite{Reuter} which was later discussed in detail in \cite{Reuter2,Percacci}. In the year 2000, in\cite{BonanoReuter} the possible effects of this approach on the Schwarzschild black hole, where the general Newton's constant was replaced by the running gravitational constant, were investigated. This type of renormalization group-improved black holes are also known as ``quantum improved (corrected) black holes". The metric for a static spherically symmetric black hole  reads
\begin{equation}\label{1.1}
ds^2=-f(r)dt^2+f(r)^{-1}dr^2+r^2d\theta^2+r^2\sin^2\theta d\phi^2
\end{equation}
where $f(r)$ is called the lapse function of the black hole with $f(r)=1-\frac{2G_0M}{r}$ for a Schwarzschild black hole in the natural units with mass of the black hole given by $M$. For a quantum-corrected Schwarzschild black hole, Newton's gravitational constant satisfies a flow equation where the gravitational constant depends on a cut-off scale. The cut-off scale can then be related to the radial distance from $r=0$ point of the black hole and takes the form \cite{BonanoReuter}
\begin{equation}\label{1.2}
G(r)=\frac{G_{0} }{1+\frac{\tilde{\omega} G_0}{r^2}}
\end{equation}
where $\tilde{\omega}$ is a positive constant that involves the quantum gravity correction and $ G_{0}$ is the general Newton's constant. The discussion on the positiveness of the constant $\tilde{\omega}$ can be found in \cite{IOYamaguchi}. Later on, such quantum correction was obtained in case of evaporating \cite{BonanoReuter2} as well as rotating black holes \cite{ReuterTurian}. There are several other analyses involving the quantum improved black holes considering the quantum effects of only the gravitational field \cite{FLRaghuraman,SausKoch,
SausKoch2,BonanoKochPlatania,
StockPawlowski,Platania,BonanoCasadioPlatania}. The effects of matter fields in the evolution of cosmological constant and Newton's gravitational constant have been studied in \cite{Percacci2,Pawlowski2,EichhornLippoldt,IOYamaguchi,ChristiansenEichhorn}.
In our analysis we shall be more focussed on the quantum corrected Reissner-Nordstr\"{o}m black hole. Even within the classical regime, as Reissner-Nordstrom metric exhibits, the electromagnetic interaction significantly modifies the Schwarzschild solution. Recently, \cite{IOYamaguchi} considered the quantum effects of the electromagnetic field along with the gravitational field and have observed how this affects the spacetime structure in the asymptotic safety scenario\cite{IOYamaguchi,HarstReuter}. It is observed that both the gravitational constant as well as the charge of the black hole flows with a cut-off scale. Another analysis \cite{RuizTurian} has recently revealed that it is sufficient to work with the flow of the Newtonian gravitational constant only keeping the charge of the black hole. The black hole metric in this case takes the form
\begin{equation}\label{1.3}
f(r)=1-\frac{2G(r)M}{r}+\frac{e^{2}G(r)}{r^{2}}
\end{equation}
where $e$ is the charge of the black hole and $G(r)$ is the running gravitational constant given in eq.(\ref{1.2}). We have in our calculation kept the charge of the black hole to be constant.

\noindent Recently there have been several analyses of a two-level atom falling into the event horizon of a black hole geometry. The initial analysis was first done in \cite{Scullyetal} where a stream of two-level atoms was considered to fall freely in the event horizon of a Schwarzschild black hole. It was observed that the atom emits virtual scalar photons and an entropy similar to that of the Bekenstein-Hawking black hole entropy was observed \cite{Hawking,Hawking2,Hawking3,Bekenstein,Bekenstein2}. This entropy was termed as the ``\textit{horizon brightened acceleration radiation}" entropy or the HBAR entropy. Later numerous amount of analyses have been made exploring several other black hole geometries \cite{Ordonez,Ordonez2,Ordonez,Ordonez3,Ordonez4,OTM,OTM2,
OTM3,OTM4}. In \cite{KaulMajumdar} logarithmic correction was observed for a quantum gravity corrected for a black hole. Later, in \cite{OTM}, we observed logarithmic correction to the HBAR entropy for a quantum-corrected black hole. Our aim in this work is to observe whether the logarithmic corrections are common to such asymptotic safety-improved black hole geometries and obtain any higher-order corrections if possible.

\noindent The paper is organized as follows. In section(\ref{S2}), we discuss the running couplings and the modified metric form for a quantum-improved charged black hole. In section (\ref{S3}), we calculate the atomic trajectories as well as the excitation and absorption probabilities for the two-level atom falling into the quantum-improved charged black hole geometry. In section (\ref{S4}), we obtain the HBAR entropy and in section (\ref{S5}) we investigate Wien's displacement law in this setup. Finally, in section (\ref{S6}), we summarize our results.

\section{The quantum improved charged black hole metric and the horizon}\label{S2}
\subsection{Running Couplings}

\noindent The theoretical way of incorporating nonperturbative quantum corrections to the solution of the non-linear Einstein field equations is by making use of the exact renormalization group flow equation. The backbone of this idea is that Einstein's general relativity is a theory at low energies coming from a theory that is valid at very high energy scales \cite{Reuter}. The improvement to Einstein's gravity is generally done in three different ways \cite{Reuter,Reuter2,ReuterTurian,
BonanoReuter,BonanoReuter2,Percacci}. The first way is to substitute the running couplings directly into the components of the metric tensor. The second way to improve is done directly to the field equations and the third way of improvement is done at the level of the action. The simplest one is to incorporate the running couplings in the metric components. The motivation for this simple approach (which is taken in \cite{BonanoReuter} as well as in this paper) is as follows. In classical general relativity, the lapse function $f(r)$ has a meaning without the need of a test particle. However, in the approach in \cite{BonanoReuter}, the improvement directly at the level of the metric $f(r)$ is regarded as a way to put in the information about the dynamics of the test particle near the body of mass $M$. This is an approximation and can be regarded as an axiom in these approaches. The prescription for constructing such a quantum-improved geometry goes as \cite{BonanoReuter,StockPawlowski,
Platania2,ReuterWeyer}
\begin{enumerate}
\item From the exact renormalization group equations obtain the scale dependent coupling constants.
\item Express the renormalization group UV cut-off scale as a function of the position which comes out to be dependent on the radial coordinate in the case of spherically symmetric geometries. This scale identification is done in primarily two ways. These are
\begin{enumerate}
\item Based on scalars constructed out of curvature entities like the Ricci scalar, $R_{\alpha\beta}R^{\alpha\beta}$, or the Kretschmann scalar.
\item Based on the UV fixed point that separates a weak-coupling regime from a strong-coupling regime.
\end{enumerate}
\item The final step is to include the quantum improvements which can be done in three ways.
\begin{enumerate}
\item In the classical solution the fundamental constants are improved by the position-dependent running couplings.
\item One can also consider the generic Einstein field equations given by 
\begin{equation}\label{2.A.1}
G_{\mu\nu}=8\pi G_0T_{\mu\nu}-g_{\mu\nu}\Lambda_0 
\end{equation}
where $G_0$ is the well-known Newtonian gravoitational constant and $\Lambda_0$ is the cosmological constant. Next, $G_0$ and $\Lambda_0$ are replaced by the position-dependent running couplings in eq.(\ref{2.A.1}) and the resulting field equations read
\begin{equation}\label{2.A.2}
G_{\mu\nu}=8\pi G(x)T_{\mu\nu}-g_{\mu\nu}\Lambda(x)~. 
\end{equation}
Here, the running of matter couplings is ignored.
\item The final way is to replace the constants by position-dependent running couplings at the action level. In the case of the Einstein-Hilbert action the improved Lagrangian density reads
\begin{equation}\label{2.A.3}
\mathcal{L}=\frac{\sqrt{-g}}{16\pi G(x)}(R-2\Lambda(x)).
\end{equation}
\end{enumerate}
The simplest approach as described earlier is to modify at the solution level.
\end{enumerate}
\noindent For a Reissner-Norsdtr\"{o}m black hole, one can consider the running of the Newtonian gravitational constant as well as the charge of the black hole. Making use of the functional renormalization group in $U(1)$-coupled gravity theories, the $U(1)$ gauge completion of $U(1)$ gauge theory was done in \cite{EichhornVersteeg}. Later this method was adopted in \cite{IOYamaguchi}, to obtain a renormalization group improved Reissner-Nordstr\"{o}m black hole where both the Newton's gravitonal constant and the charge were considered as running couplings. One can write down the beta functions corresponding to the running couplings as \cite{HarstReuter,EichhornVersteeg}
\begin{align}
k\frac{d\tilde{G}}{dk}&=2\tilde{G}\left(1-\frac{\tilde{G}}{4\pi\tilde{\alpha}}\right)\label{2.1.1}\\
k\frac{de}{dk}&=\frac{e}{4\pi}\left(\frac{be^2}{4\pi^2}-\tilde{G}\right)\label{2.1.2}
\end{align}
where $\tilde{G}(k)=k^2G(k)$, and $\tilde{\alpha}$ and $b$ are parameters. these parameters, specify the fixed points $\tilde{G}_*=4\pi\tilde{\alpha}$ and $e_*^2=(4\pi)^2\frac{\tilde{\alpha}}{b}$. As can be seen from \cite{RuizTurian}, one can also progress with a flowing Newton's gravitational constant with a fixed charge of the black hole. For such a case, the parameter $\tilde{\alpha}$ can be identified with some parameter $\tilde{\omega}'$ as
\begin{equation}\label{2.1.3}
\tilde{\omega}'=\frac{1}{4\pi\tilde{\alpha}}.
\end{equation}
This $\tilde{\omega}'$ parameter is very small and for a zero value of this parameter, the black hole spacetime loses all of the quantum-gravity contributions \cite{Reuter}. The value of $\tilde{\alpha}$ can be determined using the renormalization group equations for $U(1)$ gauge couplings \cite{EichhornVersteeg} and turns out to be positive. Hence, $\tilde{\omega}'$ is also positive. 

\noindent Solving, eq.(\ref{2.1.1}), one can then obtain
\begin{equation}\label{2.1.4}
G(k)=\frac{G(k_0)}{1+\frac{(k^2-k_0^2)}{4\pi\tilde{\alpha}}G(k_0)}
\end{equation}
where the integration has been done from $k_0$ to $k$. One can set $k_0=0$ and making use of eq.(\ref{2.1.3}), one arrives at the following result
\begin{equation}\label{2.1.5}
G(k)=\frac{G_0}{1+\tilde{\omega}'k^2G_0}
\end{equation}
where $G_0\equiv G(0)$. The cut-off identification $k$ for a spherically symmetric spacetime can be expressed in terms of a radial distance $d(r)$ as \cite{HarstReuter,EichhornVersteeg}
\begin{equation}\label{2.1.6}
k=\frac{\xi}{d(r)}.
\end{equation}
We are probing the near horizon structure of the black hole $r\gg 0$ which allows one to approximate $d(r)\sim r$. As a result the cut-off parameter $k$ goes as \cite{EichhornVersteeg}
\begin{equation}\label{2.1.7}
k\simeq\frac{\xi}{r}.
\end{equation}
Using the redefinition $\tilde{\omega}=\xi^2\tilde{\omega}'$, one can recast eq.(\ref{2.1.5}) as
\begin{equation}\label{2.1.8}
G(r)=\frac{G_0}{1+\frac{\tilde{\omega}G_0}{r^2}}
\end{equation}
which is the same as the running gravitational constant given in eq.(\ref{1.2}). 





\subsection{Black hole spacetime}
\noindent The most general renormalization group improved Reissner-Nordstr\"{o}m black solution must have a flowing gravitational constant as well as a flowing charge where the cut-off parameter is $k$. In our scenario, we restrict ourselves to flowing only the gravitational constant, and the corresponding metric structure is given in eq.(\ref{1.3}).  The justification for flowing only the gravitational constant has been provided in \cite{RuizTurian}, where it has been argued that the running gauge couplings become significant near the Planckian length scales. In our analysis, our goal is focused on the acceleration radiation of a single atom in the outskirts of the outer event horizon radius of the black hole. It therefore allows one to consider only the flow of Newton's gravitational constant. This is consistent with the flow equations (eq.(s)(\ref{2.1.1},\ref{2.1.2})).

\noindent If we need to know the trajectories of a particle near the black hole horizon, we should find the event horizon first. To find the horizons for the quantum-improved charged black hole, we need to set $ f(r) = 0$ and obtain the $r$ value(s) satisfying this equation. The horizon radii are obtained as
\begin{equation}\label{2.2.9}
r_{\pm} = MG_{0}\pm \sqrt{M^{2}G_{0}^{2}-(e^{2}+\tilde{\omega})G_{0}}~.
\end{equation}
It is important to note that $\tilde{\omega}$ is a small parameter and as a result, we can proceed with a perturbative approach. In eq.(\ref{2.2.9}), $ r_{+} $ and $ r_{-} $ are the outer (event horizon) and the inner (Cauchy horizon) horizon respectively. Since, we are mostly interested in the outer horizon, so by setting $ 2G_{0}M = 1$ and $\tilde{\omega}$ to be very small, we get the outer event horizon radius up to first-order in $\tilde{\omega}$ as
\begin{equation}
r_{+} \simeq \frac{1}{2}(1+q) - \frac{\tilde{\omega}}{2Mq}
\end{equation}
where $ q\equiv \sqrt{1-\frac{2e^{2}}{M}} $.
Our next aim is to obtain the atom trajectories while the atom is freely falling into the event horizon of the black hole.
\section{Atom falling into the quantum corrected charged black hole}\label{S3}
\noindent In \cite{Scullyetal}, a two-level atom with an angular frequency $\omega$ was considered to fall freely into the event horizon of a Schwarzschild black hole. In this work, we repeat this thought experiment in the presence of the quantum improved non-rotating charged blackhole geometry with the mass and the charge of the black hole given by $M$ and $e$. The atom is falling along a radial trajectory from infinity with zero initial velocity. The atom trajectory is given by the following set of equations for a specific black hole lapse function $f(r)$ as
\begin{equation}\label{3.1}
\tau(r)=-\int\frac{dr}{\sqrt{1-f(r)}},~t(r)=-\int \frac{dr}{f(r)\sqrt{1-f(r)}}~.
\end{equation}
To obtain the trajectories in the quantum-improved charged black hole geometry, we consider a change of variables given by $z\equiv r-r_+$ such that the value of the parameter $z$ is zero at the event horizon of the black hole. As we are considering near horizon regimes, $z\ll1$. Making use of a new constant as $a\equiv\frac{1}{2}(1+q)$ and using a perturbative approach, we obtain the atom trajectories as
\begin{align}
\tau(r)=&-\int \frac{dr}{\sqrt{1-f(r)}}\nonumber\\
\simeq& -\int dz\left(1+\frac{2qy}{(q+1)^2}-\frac{2\tilde{\omega}y(q^2+1)}{Mq(q+1)^4}\right)\nonumber\\
=& -z-\frac{qz^2}{(q+1)^2}\left(1-\frac{\tilde{\omega}(q^2+1)}{Mq^2(q+1)^2}\right) +C_{1} ~\label{3.2}
\\
t(r)=&-\int\frac{dr}{f(r)\sqrt{1-f(r)}}\nonumber\\
&\simeq -\left(\frac{(1+q)^{2}}{4q} +\frac{(1+q^2)\tilde{\omega}}{4Mq^3}\right) \ln z \nonumber\\&+ \left(\frac{(1-q)^2}{4q^2}-\frac{3}{2}+\frac{\tilde{\omega}(1-q+q^2)}{2Mq^4}\right)z+C_{2}\label{3.3}
\end{align}
with $C_{1}$ and $C_{2}$ being the integration constants\footnote{It is important to note that the form of the proper time $\tau(z)$ in terms of $z$ is kept up to second order in $z$. It is necessary as later in the calculation we need to calculate $\frac{\partial\tau(z)}{\partial z}$ which will give a proper estimate of the transition probability and a correct perturbative result when $\tau(z)$ is kept upto second order in $z$.}. The constant $q$ as has been defined earlier is related to the charge of the black hole.  For the sake of calculation, we have set $2G_{0}M$ to be unity which will be restored later. Now for a scalar photon in the Regge-Wheeler coordinates, we obtain
\begin{equation}\label{3.4}
\begin{split}
r_*(r)=&\int\frac{d(r)}{f(r)}\\
\implies r_*(z)\simeq&\left(\frac{(1+q)^2}{4q}+\frac{\tilde{\omega}(q^2+1)}{4Mq^3}\right)\ln z\nonumber\\&-\left(\frac{(q+1)^2}{4q^2}-\frac{\tilde{\omega}}{2Mq^3}\right)z+C_3
\end{split}
\end{equation}
with $C_3$ being an integration constant.
The covariant Klein-Gordon equation takes the form
\begin{equation}\label{3.5}
\begin{split}
\frac{1}{\sqrt{-g}}\partial_\mu\left(\sqrt{-g}g^{\mu\nu}\partial_\nu\right)\psi(t,\vec{x})-m^2\psi(t,\vec{x})=0~.
\end{split}
\end{equation}
For a massless scalar photon with  wavefunction $\psi$ and neglecting the contribution from the angular part, one can recast eq.(\ref{3.5}) as
\begin{equation}\label{3.6}
\frac{1}{T(t)}\frac{d^{2}T(t)}{dt^{2}}-\frac{f(r)}{r^{2}R(r)}\frac{d}{dr}(r^{2}f(r)\frac{dR(r)}{dr}) = 0
\end{equation}
where the $\psi(t,r)$ has the form
\begin{equation}\label{3.7}
\psi (r,t) = R(r)T(t)~.
\end{equation}
The general solution of eq.(\ref{3.6}) has the form
\begin{align}\label{3.8}
\psi_\nu(t,r)=\frac{\Psi_\nu(t,r)}{r}
\end{align}
where the analytical form of $\Psi_\nu(t,r)$ reads
\begin{equation}\label{3.9}
\Psi_\nu(t,r)=\exp\left[- i\nu t+ i\nu\int\frac{dr}{f(r)}\right]~.
\end{equation}
In the above equation, $\nu$ denotes the frequency of the emitted photon measured by a distant observer.
It is important to note that since we are mainly interested in the near horizon behaviour of the solution, hence the spherically symmetric solution can be easily approximated by \cite{Scullyetal} $\psi_\nu(t,r)\sim\frac{\Psi_\nu(t,r)}{r_+}$. It is important to note that $r_+\sim1$ for $2G_0M=1$, and as a result it is always possible to approximate the solution of the single-mode massless scalar field using $\Psi_\nu(t,r)$ only.

\noindent With the atom-field trajectories in hand, we shall now proceed to calculate the transition probability. To calculate the transition probability, we need to write down the atom-field interaction Hamiltonian which is given as
\begin{equation}\label{3.10}
\hat{V}_I(\tau)=\hbar \mathcal{G}[\hat{b}_\nu \psi_\nu(t(\tau),r(\tau))+H.c.][\hat{\zeta}e^{-i\Omega\tau}+h.c.]~.
\end{equation}
where $\mathcal{G}$ is the atom-field coupling constant, $\hat{b}_\nu$ is the annihilation operator corresponding to the scalar photon, $\Omega$ is the transition frequency corresponding to the atom and $ \hat{\zeta} = |g \rangle \langle e| $ is the lowering operator of the atom. Initially, the atom is in the ground state and there is no scalar photon in this ground state. Here the field is initially in the Boulware vacuum \cite{Boulware}.

\noindent Before the interaction starts, one can always consider the state of the system to be a tensor product state given as $|0_\nu,g\rangle$ and the final state of the system is given as $|1_\nu,e\rangle$. It is important to note that there is no photon initially in the ground state of the atom and a simultaneous emission happens with that of the excitation of the atom to its excited state. That is why it is known as a virtual transition and the corresponding atom-field transition probability is given by
\begin{equation}\label{3.11}
\begin{split}
P_{exc}&=\frac{1}{\hbar^2}\left|\int d\tau\langle 1_\nu,e|\hat{V}_{I}(\tau)|0_\nu,g\rangle\right|^2\\ 
&=\mathcal{G}^2\left|\int_{0}^{z_f}dz \left(\frac{d\tau}{dz}\right) e^{i\nu t(z)-i\nu r_*(z)}e^{i\Omega\tau(z)}\right|^2
\end{split}
\end{equation}
where the forms of $\tau(z)$ and $t(z)$ are given in eq.(s)(\ref{3.2},\ref{3.3}) and the expression for $ r_{*}(z) $ is obtained in eq.(\ref{3.4}). In the above equation, $z_f$ denotes the initial distance of a freely falling atom from the event horizon of the black hole. The form of the excitation probability from eq.(\ref{3.11}) now can be recast as
\begin{widetext}
\begin{align}
P_{\text{exc}} &= \mathcal{G}^{2}\biggr|\int_{0}^{z_f}dz \left[1+\frac{2qz}{(1+q)^{2}}\left(1-\frac{\tilde{\omega}(1+q^2)}{Mq^2(1+q)^2}\right)\right]e^{-i\nu \left[ \frac{(1+q)^{2}}{2q} +\frac{\tilde{\omega}(1+q^2)}{2Mq^{3}}\right] \ln {z}+i\nu\left(\frac{1-2q^2}{2q^2}+\frac{\tilde{\omega}(1-q)^2}{2Mq^4}\right)z}e^{-i\Omega z}\biggr|^{2}\nonumber\\
&=\mathcal{G}^{2}\biggr|\int_{0}^{z_f}dz \left[1+\frac{2qz}{(1+q)^{2}}\left(1-\frac{\tilde{\omega}(1+q^2)}{Mq^2(1+q)^2}\right)\right]z^{-i\nu \left[ \frac{(1+q)^{2}}{2q} +\frac{\tilde{\omega}(1+q^2)}{2Mq^{3}}\right]}e^{-i\Omega z\left(1-
\frac{\nu}{\Omega}\left(\frac{(1-2q^2)}{2q^2}+\frac{\tilde{\omega}(1-q)^2}{2Mq^4}\right)\right)}\biggr|^{2}~.
 \label{3.12}
\end{align}
\end{widetext}
Now, let us take a change of variable $ z\Omega = y $, where $ \Omega\gg 1 $ and hence we obtain the form of the excitation probability $ P_{\text{exc}} $ as
\begin{equation}
\begin{split}
P_{\text{exc}}&=\frac{\mathcal{G}^{2}}{\Omega^2}\biggr|\int_{0}^{y_f}dy \left[1+\frac{2qy}{\Omega(1+q)^{2}}\left(1-\frac{\tilde{\omega}(1+q^2)}{Mq^2(1+q)^2}\right)\right]\\&\times y^{-i\nu \left[ \frac{(1+q)^{2}}{2q} +\frac{\tilde{\omega}(1+q^2)}{2Mq^{3}}\right]}e^{-iy\left[1-
\frac{\nu}{\Omega}\left(\frac{(1-2q^2)}{2q^2}+\frac{\tilde{\omega}(1-q)^2}{2Mq^4}\right)\right]}\biggr|^{2}\label{3.13}
\end{split}
\end{equation}
where $y_f=z_f\Omega$.
In order to execute the above integral, we now consider another change of variable given as
\begin{equation}\label{3.14}
\xi\equiv y\left[1+
\frac{\nu}{\Omega}\left(1-\frac{1}{2q^2}-\frac{\tilde{\omega}(1-q)^2}{2Mq^4}\right)\right]
\end{equation}
such that the integration limits remain the same. Using the change of variables in the above equation, we can recast the transition probability in eq.(\ref{3.13}) as
\begin{widetext}
\begin{equation}
\begin{split}
P_{\text{exc}} &\simeq \frac{\mathcal{G}^{2}}{\Omega^{2}}\left[ 1-\frac{2\nu}{\Omega}\left(1-\frac{1}{2q^2}-\frac{\tilde{\omega}(1-q)^2}{2Mq^4}\right)\right] \biggr|\int_0^{\xi_f} d\xi \biggr[1+\frac{2q\xi}{\Omega (1+q)^{2}}\biggr[1-\frac{\nu}{\Omega}\left(1-\frac{1}{2q^2}\right)-\frac{\tilde{\omega}(1+q^2)}{Mq^2(1+q)^2}\\&-\frac{\tilde{\omega}\nu(1-3q^2)}{2Mq^2\Omega(1+q)^2}\biggr]\biggr] 
\times \xi^{-2i\nu \left[\frac{(1+q)^{2}}{4q}+\frac{\tilde{\omega}(1+q^2)}{4Mq^{3}}\right]} \exp[-i\xi]\biggr|^{2} \label{3.15}
\end{split}
\end{equation}
\end{widetext}
where $\xi_f=y_f\left[1+
\frac{\nu}{\Omega}\left(1-\frac{1}{2q^2}-\frac{\tilde{\omega}(1-q)^2}{2Mq^4}\right)\right]$.

\noindent We define a new quantity given as
\begin{equation}\label{3.15a}
\begin{split}
\mathcal{B}&\equiv\frac{2q}{\Omega (1+q)^{2}}\biggr[1-\frac{\nu}{\Omega}\left(1-\frac{1}{2q^2}\right)-\frac{\tilde{\omega}(1+q^2)}{Mq^2(1+q)^2}\\&-\frac{\tilde{\omega}\nu(1-3q^2)}{2Mq^2\Omega(1+q)^2}\biggr]~.
\end{split}
\end{equation}
Using eq.(\ref{3.15a}), we can recast eq.(\ref{3.15}) as
\begin{equation}\label{3.15b}
\begin{split}
P_{\text{exc}} \simeq& \frac{\mathcal{G}^{2}}{\Omega^{2}}\left[ 1-\frac{2\nu}{\Omega}\left(1-\frac{1}{2q^2}-\frac{\tilde{\omega}(1-q)^2}{2Mq^4}\right)\right] \biggr|\int_0^{\xi_f} d\xi \\\times&\left(1+\mathcal{B}\xi\right) 
\xi^{-2i\nu \left[\frac{(1+q)^{2}}{4q}+\frac{\tilde{\omega}(1+q^2)}{4Mq^{3}}\right]} \exp[-i\xi]\biggr|^{2}~.
\end{split}
\end{equation}
We need to solve the above integral, in order to obtain the form of the transition probability from the above equation. If we define $\lambda\equiv\frac{(1+q)^{2}}{4q}+\frac{\tilde{\omega}(1+q^2)}{4Mq^{3}}$ then we need to solve two integrals given as
\begin{align}
I_1&=\int_0^{\xi_f}d\xi ~\xi^{-2i\nu\lambda} e^{-i\xi}\label{I1}\\
I_2&=\int_0^{\xi_f}d\xi~ \xi^{1-2i\nu\lambda} e^{-i\xi}~.\label{I2}
\end{align}
As has been discussed in \cite{Ordonez,OTM3}, the integration limits in $I_1$ can be extended from $\xi_f$ to $\infty$. The primary reason behind this consideration is that while $e^{-i\xi}$ behaves as a highly oscillatory function, $\xi^{-2i\nu\lambda}$ has a ``Russian doll"-like behaviour resulting in decreasing negligible contributions towards the overall integration value with higher values of $y_f$ (as a result $\xi_f$). Whereas the $I_2$ integral in eq.(\ref{I2}) has no such properties resulting in a definite integral with finite integration limits. As, we have already observed in \cite{OTM3}, the integral $I_2$ gives incomplete Gamma functions which leads to a slightly deformed Planckian behaviour. However, if one is considering the atom very near the event horizon or $z_f$ very close to zero then the Planckian nature is much more prominent as we shall see while calculating the form of the transition probability. As $\mathcal{B}\xi\ll1$, we can simply ignore the effects from $I_2$ and consider $I_1$ instead. Making use of the following result

\begin{equation}\label{3.16}
\int_{0}^{\infty} dx~ x^{2i\nu} \exp[ix] = \frac{\pi \exp[-\pi \nu]}{\sinh{(2\pi \nu)}\Gamma[-2i\nu]}
\end{equation}
we obtain the analytical form of the excitation probability as
\begin{widetext}
\begin{align}
P_{\text{exc}} &\simeq\frac{4\pi \mathcal{G}^{2}\nu}{\Omega^{2}}\left[\frac{(1+q)^{2}}{4q}\left[1-\frac{2\nu}{\Omega}\left(1-\frac{1}{2q^2}\right)\right]+\frac{\tilde{\omega}(1+q^2)}{4Mq^3}\left[1-\frac{\nu(q^4+3q^2-2)}{\Omega q^2(1+q^2)}\right]\right]\frac{1}{e^{4\pi \nu \left(\frac{(1+q)^{2}}{4q}+\frac{\tilde{\omega}(1+q^2)}{4Mq^3}\right)} - 1} \label{3.17}~.
\end{align}
\end{widetext}
Eq.(\ref{3.17}) is one of the main results in our paper. For the sake of completeness of our calculation, we shall also keep the $I_2$ integral in our calculation as well and obtain the transition probability\footnote{We shall although be using eq.(\ref{3.17}) for obtaining the rest of the results in our paper.}. The transition probability in eq.(\ref{3.15b}), can simply be written as a combination of two integrals
\begin{equation}\label{3.17a}
\begin{split}
P_{\text{exc}}&=\frac{\mathcal{A}\mathcal{G}^2}{\Omega^2}\left(I_1\rvert_{\xi_f\rightarrow\infty}+\mathcal{B}I_2\right)\left(I_1^*\rvert_{\xi_f\rightarrow\infty}+\mathcal{B}I_2^*\right)
\end{split}
\end{equation}
where
\begin{equation}\label{3.17b}
\mathcal{A}\equiv 1-\frac{2\nu}{\Omega}\left(1-\frac{1}{2q^2}-\frac{\tilde{\omega}(1-q)^2}{2Mq^4}\right)~.
\end{equation}
Denoting $I_1\rvert_{\xi_f\rightarrow\infty}$ as $I_{1\infty}$, we obtain the form of the combined integral as
\begin{equation}\label{3.17c}
\begin{split}
&I_{1\infty}+\mathcal{B}I_2\\&=-2\lambda\nu e^{-\pi\nu\lambda}\left(\Gamma[-2i\lambda\nu]+\frac{\mathcal{B}}{2\lambda\nu}\gamma\left[2-2i\lambda\nu,i\xi_f\right]\right)
\end{split}
\end{equation}
where $\gamma\left[2-2i\lambda\nu,i\xi_f\right]$ is the lower incomplete gamma function given as
\begin{equation}\label{3.17d}
\begin{split}
\gamma\left[2-2i\lambda\nu,i\xi_f\right]
=\Gamma[2-2i\lambda\nu]-\Gamma\left[2-2i\lambda\nu,i\xi_f\right]~.
\end{split}
\end{equation}
In eq.(\ref{3.17}), the $I_{1\infty}I^*_{1\infty}$ term only contributed from eq.(\ref{3.17a}). It is easy to check from this deformed transition probability that it is non-Planckian in nature but we restrict ourselves from investigating that part in our current analysis. We confine ourselves strictly to a near-horizon analysis. We shall now move towards a direct comparison of the above black hole metric with that of the general Reissner-Norsdstr\"{o}m black hole metric. 
The transition probability in eq.(\ref{3.17}) can more simply be expressed as (considering the $\nu\ll\Omega$ approximation) 
\begin{widetext}
\begin{equation}\label{3.17e}
\begin{split}
P_{\text{exc}} &\simeq\frac{4\pi \mathcal{G}^{2}\nu}{\Omega^{2}}\left[\frac{(1+q)^{2}}{4q}+\frac{\tilde{\omega}(1+q^2)}{4Mq^3}\right]\frac{1}{e^{4\pi \nu \left(\frac{(1+q)^{2}}{4q}+\frac{\tilde{\omega}(1+q^2)}{4Mq^3}\right)} - 1}~.
\end{split}
\end{equation}
\end{widetext}
In order to truly understand the importance of the above result, we start by setting $\tilde{\omega}\rightarrow 0$ limit in eq.(\ref{3.17e}) which surely reproduces the Reissner-Nordstr\"{o}m case without any quantum-gravity correction. 
In the $e^2\ll M$ limit, the transition probability takes the form given as (neglecting any higher order $\nu/\Omega$ contribution)
\begin{equation}\label{3.18}
\begin{split}
P_{\text{exc}}\bigr\rvert_{\tilde{\omega}\rightarrow 0} &\simeq\frac{4\pi \mathcal{G}^{2}\nu}{\Omega^{2}}\left(1+\frac{e^4}{4M^2}\right) \frac{1}{e^{4\pi \nu \left(1+\frac{e^4}{4M^2}\right)} - 1}~.
\end{split}
\end{equation}
It is crucial to note that all lower order $\frac{e^2}{M}$ kind of contributions have cancelled out and the charge contribution appears in the next higher order in the coefficient factor as well as the exponent. Such lower order cancellation also happens for extra-dimensional black holes with a tidal charge \cite{OTM4}. Now, in the presence of a flowing gravitational constant, the same transition probability takes the form (again neglecting $\frac{\nu}{\Omega}$ kind of small contributions as $\nu\ll\Omega$)
\begin{widetext}
\begin{equation}\label{3.19}
\begin{split}
P_{\text{exc}}&\simeq\frac{4\pi \mathcal{G}^{2}\nu}{\Omega^{2}}\left(1+\frac{\tilde{\omega}}{2M}+\frac{\tilde{\omega}e^2}{M^2}+\frac{e^4}{4M^2}\left(1+\frac{9\tilde{\omega}}{M}\right)\right)\frac{1}{e^{4\pi \nu \left(1+\frac{\tilde{\omega}}{2M}+\frac{\tilde{\omega}e^2}{M^2}+\frac{e^4}{4M^2}\left(1+\frac{9\tilde{\omega}}{M}\right)\right)} - 1}~.
\end{split}
\end{equation}
\end{widetext}
We observe from the above result that $e^2/M$ order of contribution occurs in the transition probability purely due to the induced quantum gravitational effects into the theory in contrast to the result obtained in eq.(\ref{3.18}). With a proper dimensional reconstruction, we can now recast eq.(\ref{3.17}) in the following form
\begin{widetext}
\begin{align}
P_{\text{exc}} &\simeq\frac{4\pi \mathcal{G}^{2}\nu}{\Omega^{2}}\left[\frac{2G_{0}M}{c^{3}}\frac{(1+q)^{2}}{4q}\left[1-\frac{2\nu}{\Omega}\left(1-\frac{1}{2q^2}\right)\right]+\frac{\hbar\tilde{\omega}(1+q^2)}{4Mc^2q^3}\left[1-\frac{\nu(q^4+3q^2-2)}{\Omega q^2(1+q^2)}\right]\right]\nonumber\\\times&\frac{1}{e^{4\pi \nu \left(\frac{2G_{0}M}{c^{3}}\frac{(1+q)^{2}}{4q}+\frac{\hbar\tilde{\omega}(1+q^2)}{4Mc^2q^3}\right)} - 1}~. \label{3.20}
\end{align}
Now, the photon absorption probability can be directly obtained by replacing $\nu$ by $ -\nu$ in the above equation, and the analytical form is given by
\begin{align}
P_{\text{abs}} &\simeq\frac{4\pi \mathcal{G}^{2}\nu}{\Omega^{2}}\left[\frac{2G_{0}M}{c^{3}}\frac{(1+q)^{2}}{4q}\left[1+\frac{2\nu}{\Omega}\left(1-\frac{1}{2q^2}\right)\right]+\frac{\hbar\tilde{\omega}(1+q^2)}{4Mc^2q^3}\left[1+\frac{\nu(q^4+3q^2-2)}{\Omega q^2(1+q^2)}\right]\right]\nonumber\\\times&\frac{1}{1-e^{-4\pi \nu \left(\frac{2G_{0}M}{c^{3}}\frac{(1+q)^{2}}{4q}+\frac{\hbar\tilde{\omega}(1+q^2)}{4Mc^2q^3}\right)}}~. \label{3.21}
\end{align}
\end{widetext}
In these types of calculations, one can consider the atomic frequency to be much higher than the field frequency, i.e. $ \Omega \gg \nu $. In that case, the excitation probability in eq.(s)(\ref{3.20}) takes the form given as \footnote{An alternative way of obtaining the transition probability will be to follow the near and beyond near horizon methods followed in \cite{Ordonez,OTM3}.}
\begin{align}
P_{\text{exc}} &\simeq\frac{4\pi \mathcal{G}^{2}\nu}{\Omega^{2}}\left[\frac{2G_{0}M}{c^{3}}\frac{(1+q)^{2}}{4q}+\frac{\hbar\tilde{\omega}(1+q^2)}{4Mc^2q^3}\right]\nonumber\\&\times\frac{1}{\exp\left[4\pi \nu \left(\frac{2G_{0}M}{c^{3}}\frac{(1+q)^{2}}{4q}+\frac{\hbar\tilde{\omega}(1+q^2)}{4Mc^2q^3}\right)\right] - 1}~.\label{3.22}
\end{align}
and the absorption probability in eq.(\ref{3.21}) reads
\begin{align}
P_{\text{abs}} &\simeq\frac{4\pi \mathcal{G}^{2}\nu}{\Omega^{2}}\left[\frac{2G_{0}M}{c^{3}}\frac{(1+q)^{2}}{4q}+\frac{\hbar\tilde{\omega}(1+q^2)}{4Mc^2q^3}\right]\nonumber\\&\times\frac{1}{1-\exp\left[-4\pi \nu \left(\frac{2G_{0}M}{c^{3}}\frac{(1+q)^{2}}{4q}+\frac{\hbar\tilde{\omega}(1+q^2)}{4Mc^2q^3}\right)\right]}~.
\end{align}
The excitation and absorption probability contain the Planck-like factor containing the field frequency whereas the atomic frequency appears in the coefficient of the transition probabilities.
\section{Calculating the modified HBAR entropy}\label{S4}
\noindent The HBAR entropy (horizon brightened acceleration radiation entropy) was introduced to distinguish between the Bekenstein Hawking entropy and entropy due to the atom falling into the black hole \cite{Scullyetal}. In this section, we will calculate the HBAR entropy for the quantum-improved charged black hole. Here we have considered a cloud of two-level atoms with angular frequency $\Omega$ and we consider that they are falling in a linear stream into the quantum-improved charged black hole with an infall rate of $\kappa$ and the atoms emit and absorb the acceleration radiation. We shall calculate the entropy in this case by using the method of quantum statistical mechanics for which we need to find the density matrix for the field first. 

\noindent For a microscopic change of $\delta \rho^{i}$ of the field density matrix due to one atom, the total change in the same for $\Delta \mathcal{N}$ number of atoms is given by 
\begin{equation}\label{4.1}
\Delta \rho=\sum\limits_j \delta\rho_j=\Delta \mathcal{N}\delta\rho
\end{equation} 
where for a time duration of $\Delta t$, we know that
\begin{equation}\label{4.2}
\frac{\Delta \mathcal{N}}{\Delta t}=\kappa~.
\end{equation}
Putting the form of $\Delta\mathcal{N}$ from eq.(\ref{4.1}) in the above equation, we obtain
\begin{equation}\label{4.3}
\frac{\Delta \rho}{\Delta t}=\kappa\delta\rho~.
\end{equation}
If $|n\rangle$ denotes the state of the field containing $n$ number of scalar photons, then the equation of motion for the field density matrix can be written as
\begin{equation}\label{4.4}
\begin{split}
\dot{\rho}_{n,n}=&-\Gamma_{\text{abs}}\left(n\rho_{n,n}-(n+1)\rho_{n+1,n+1}\right)\\&-\Gamma_{\text{exc}}\left((n+1)\rho_{n,n}-n\rho_{n-1,n-1}\right)
\end{split}
\end{equation}
where $\Gamma_{\text{exc}}$ and $\Gamma_{\text{abs}}$ are the excitation rate and absorption rate defined as $ \Gamma_{\text{exc}/\text{abs}} = \kappa P_{\text{exc}/\text{abs}}$. We need the steady-state solution to find the HBAR entropy. For the steady state solution, we shall set $ \ddot{\rho}_{n,n} = 0 $ when $ n = 0 $, and hence we shall obtain the following relation by using $\text{tr}[\rho]=1$

\begin{equation}\label{4.5}
\rho_{n,n}^{s.s.} = \left(\frac{\Gamma_{\text{exc}}}{\Gamma_{\text{abs}}}\right)^{n}\left(1-\frac{\Gamma_{\text{exc}}}{\Gamma_{\text{abs}}}\right)~.
\end{equation}
\begin{widetext}
In the high atomic frequency limit ($\nu\ll\Omega$) and making  use of eq.(s)(\ref{3.20},\ref{3.21}), we obtain
\begin{align}
\frac{\Gamma_{exc}}{\Gamma_{abs}} &\simeq e^{-4\pi\nu \left[\frac{2G_{0}M}{c^{3}}\frac{(1+q)^{2}}{4q}+\frac{\hbar\tilde{\omega}(1+q^2)}{4Mc^2q^3}\right]} \left[1-\frac{4\nu}{\Omega}\left(1-\frac{1}{2q^2}\right)-\frac{\nu}{\Omega}\frac{\hbar \tilde{\omega}c(q^4+3q^2-2)}{G_{0}M^{2}q^4(1+q)^{2}}\right]~.\label{4.6}
\end{align}
Using the above equation back in eq.(\ref{4.5}), we obtain the steady state solution of the density matrix to be
\begin{align}\label{4.7}
\rho_{n,n}^{s.s.} &= e^{-4\pi n\nu \left[\frac{2G_{0}M}{c^{3}}\frac{(1+q)^{2}}{4q}+\frac{\hbar\tilde{\omega}(1+q^2)}{4Mc^2q^3}\right]} \left[1-\frac{4\nu}{\Omega}\left(1-\frac{1}{2q^2}\right)-\frac{\nu}{\Omega}\frac{\hbar \tilde{\omega}c(q^4+3q^2-2)}{G_{0}M^{2}q^4(1+q)^{2}}\right]^{n} \left(1-\frac{\Gamma_{exc}}{\Gamma_{abs}}\right)~.
\end{align}
\end{widetext}
The von-Neumann entropy for the system is given by
\begin{equation}\label{4.8}
S_\rho=-k_B\sum\limits_{n,\nu}\rho_{n,n}\ln(\rho_{n,n})
\end{equation}
and the rate of change of entropy due to the generation of real photons using the steady state solution of the field density matrix is \cite{Scullyetal}

\begin{equation}\label{4.9}
\dot{S_\rho}\simeq-k_B\sum\limits_{n,\nu}\dot{\rho}_{n,n}\ln(\rho^{s.s.}_{n,n})~.
\end{equation}
Now, putting the steady state solution of the density matrix from eq.(\ref{4.7}) in the above equation, we obtain
\begin{widetext} 
\begin{align}
\dot{S_\rho} & \simeq 4\pi K_{B} \left[\frac{\mathcal{R}}{c}\frac{(1+q)^{2}}{4q} +\frac{\hbar \tilde{\omega}(1+q^2)}{4Mc^{2}q^{3}}\right] \sum\limits_{\nu}\dot{\bar{n}}_{\nu}\nu + \frac{4K_{B}}{\Omega}\left(1-\frac{1}{2q^2}\right)\sum\limits_{\nu}\dot{\bar{n}}_{\nu}\nu+\frac{\hbar\tilde{\omega}K_{B}c(q^4+3q^2-2)}{\Omega G_{0}M^{2}q^4(1+q)^{2}}\sum\limits_{\nu}\dot{\bar{n}}_{\nu}\nu \label{4.10}
\end{align}
where $\mathcal{R}$ is the Schwarzschild radius given by $\mathcal{R}= \frac{2G_{0}M}{c^{2}} $ and  $\dot{\bar{n}}_\nu$ is the flux due to photons emitted from the atoms freely falling in the black hole and the total rate of energy loss due to emitted photons is $\hbar\sum\limits_\nu\dot{\bar{n}}_\nu\nu=\dot{m}_pc^2$. So now the rate of change of entropy becomes 
\begin{align}
\dot{S_\rho} & = \frac{4\pi K_{B}}{\hbar} \left[\frac{\mathcal{R}}{c}\frac{(1+q)^{2}}{4q} +\frac{\hbar \tilde{\omega}(1+q^2)}{4Mc^{2}q^{3}}\right] \dot{m}_pc^2+ \frac{4K_{B}}{\hbar\Omega}\left(1-\frac{1}{2q^2}\right)\dot{m}_pc^2+\frac{\tilde{\omega}K_{B}c(q^4+3q^2-2)}{\Omega G_{0}M^{2}q^4(1+q)^{2}}\dot{m}_pc^2\label{4.11}~.
\end{align}
\end{widetext}
In our current analysis, the area of the quantum-corrected charged black hole is given by
\begin{equation}\label{4.12}
\begin{split}
A_{\text{QCBH}}&=4\pi r_+^2\simeq\frac{4\pi G_{0}^{2}M^{2}}{c^{4}}(1+q)^{2} - \frac{4\pi\hbar \tilde{\omega}G_{0}}{qc^{3}}\left(1+q\right)~.
\end{split}
\end{equation}  
The rate of change of the black hole area from the above equation is obtained as
\begin{align}
\dot{A}_{\text{QCBH}} &= \frac{8\pi G_{0}^{2}M\dot{M}}{c^{4}q}(1+q)^{2} + \frac{4\pi \hbar \tilde{\omega}G_{0}}{c^{3}q}\left[\frac{1}{q^{2}}-1\right]\left(\frac{\dot{M}}{M}\right) \label{4.13}
\end{align}
where $ \dot{M} $ is the total change of mass of the black hole due to the emission of scalar photons and the addition of freely falling atoms. It is therefore always possible to express $\dot{M}$ as \cite{Scullyetal,OTM}
\begin{equation}
\dot{M}=\dot{m}_p+\dot{m}_{\text{atom}}
\end{equation}
where $\dot{m}_p$ denotes the rate of change in the mass of the black hole due to emitting photons and $\dot{m}_{\text{atom}}$ denotes the rate of change of the mass of the black hole due to the infalling atoms. Hence, we can define the rate of change of the area of the black hole due to the emission of photons as
 \begin{align}
\dot{A}_{p} &=\frac{8\pi G_{0}^{2}M\dot{m}_{p}}{c^{4}q}(1+q)^{2} + \frac{4\pi \hbar \tilde{\omega}G_{0}}{c^{3}q}\left(\frac{1}{q^{2}}-1\right)\left(\frac{\dot{m}_{p}}{M}\right) \label{4.14}~.
\end{align}
When no atom is falling into the black hole, the change in the area of the black hole due to freely falling atoms, $A_{\text{atom}}$, is zero and as a result $A_{p}$ is same as the area of the black hole $ A_{\text{QCBH}} $ \cite{Scullyetal,OTM}. Now, if we consider very high atomic frequency such that $\hbar \Omega\gg m_p c^2$, we can neglect the last two terms in eq.(\ref{4.11}), and we can write down the rate of change of the entropy (keeping terms upto $\mathcal{O}(\tilde{\omega}^2)$) as
\begin{equation}\label{4.14a}
\begin{split}
\dot{S}_\rho\simeq&\frac{4\pi K_{B}}{\hbar} \left[\frac{\mathcal{R}}{c}\frac{(1+q)^{2}}{4q} +\frac{\hbar \tilde{\omega}(1+q^2)}{4Mc^{2}q^{3}}\right] \dot{m}_pc^2\\
\simeq&\frac{k_B\dot{A}_p c^3}{4\hbar G_0}+\pi \tilde{\omega}k_B\frac{d}{dt}\left(\ln A_p\right)~.
\end{split}
\end{equation}
The above result can be written in much more dimensionally correct form as
\begin{equation}\label{4.14b}
\dot{S}_\rho\simeq\frac{d}{dt}\left[\frac{k_B c^3}{4\hbar G_0}A_p+\pi \tilde{\omega}k_B\ln \left[\frac{A_p c^3}{4\hbar G_0}\right]\right]~.
\end{equation}
This is a very strong result in our paper as it exactly matches with the HBAR entropy result in \cite{OTM} up to the sub-leading logarithmic correction term. The surprising thing is that the HBAR entropy also has matching coefficients in the subleading correction which proves the universality of logarithmic corrections for such quantum-improved black hole geometries. The leading order term as can be seen from eq.(\ref{4.14a}) is following the ``\textit{area divided four law}" observed in all of the earlier analyses \cite{Scullyetal,Ordonez4,OTM,OTM2,
OTM3,OTM4}.
Here, the HBAR entropy we are getting is the Bekenstein-Hawking entropy with some quantum gravity correction terms which contain the charge of the black hole. We can see that some of the correction terms are similar to the entropy correction terms obtained for the quantum-improved Schwarzschild black hole \cite{KaulMajumdar}. 
\section{Wien Displacement for the quantum corrected charged black hole}\label{S5}
\begin{figure}[ht!]
\begin{center}
\includegraphics[scale=0.2]{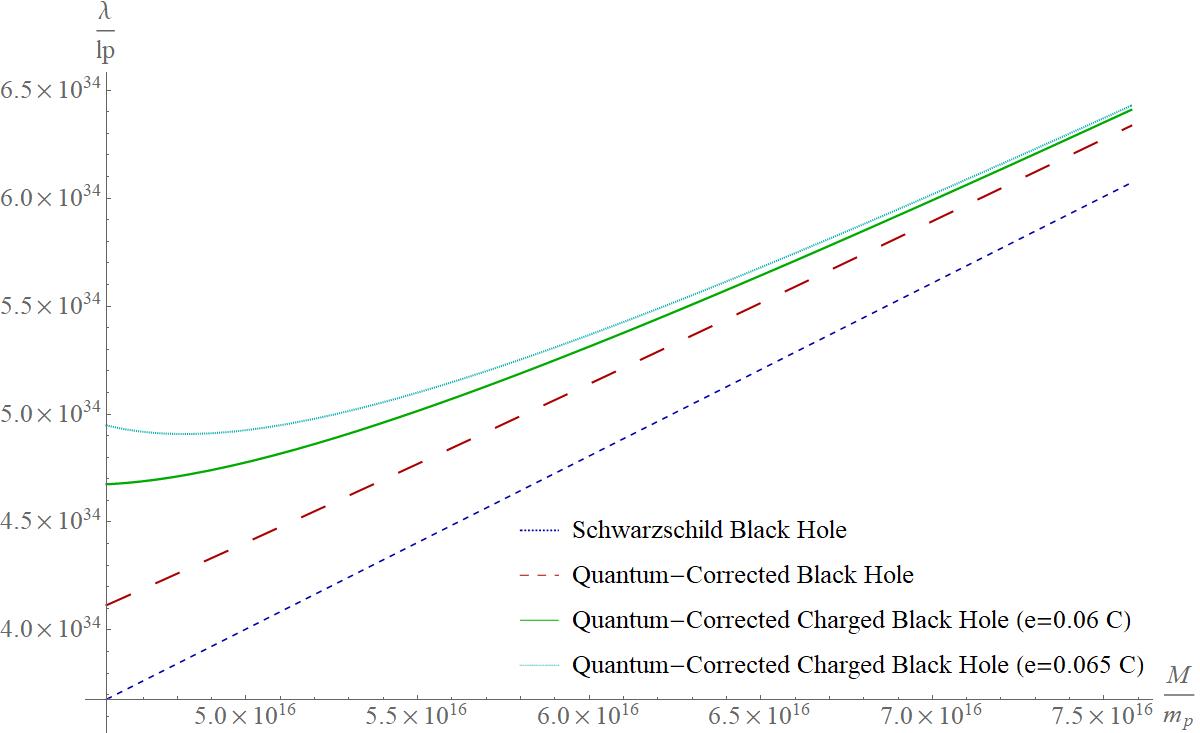}
\caption{Plot of the dimensionless wavelength ($\lambda/l_p$) from the Wien's displacement law against the dimensionless mass ($M/m_p$) of the black hole\label{Lambda}. Here $l_p$ and $m_p$ indicate the Planck length and Planck mass respectively.}
\end{center}
\end{figure}

\noindent In this section, we investigate the validity of Wien's displacement law for the background quantum corrected black hole geometry and compare the wavelength plotted against the mass as well as the charge parameter for this black hole with the classical Schwarzcihild as well as the quantum corrected Schwarzschild geometries. The excitation probabilities for a Schwarzschild geometry  \cite{Scullyetal} as well as the quantum corrected black hole geometry
\begin{align}
P_{\text{exc}}^{\text{Sch.}}(\nu)&\simeq\frac{4\pi\mathcal{G}^2\mathcal{R}\nu }{c\Omega^2}\left(1-\frac{4\nu}{\Omega}\right)\frac{1}{e^{\frac{4\pi\mathcal{R}\nu}{c}}-1}\label{5.1}\\
P_{\text{exc}}^{\text{QBH}}(\nu)&\simeq\frac{4\pi\mathcal{G}^2\nu}{\Omega^2}\left(\frac{\mathcal{R}}{c}+\frac{\hbar\tilde{\omega}}{2Mc^2}\right)\left(1-\frac{\nu}{\Omega}\right)\nonumber\\&\times\frac{1}{e^{4\pi\nu\left(\frac{\mathcal{R}}{c}+\frac{\hbar\tilde{\omega}}{2Mc^2}\right)}-1}\label{5.2}
\end{align}
where we have used eq.(\ref{3.20}) and set the $q\rightarrow 1$ ($e\rightarrow 0$) limit.
The temperatures of the thermal baths corresponding to the thermal distributions in eq.(s)(\ref{5.1},\ref{5.2}) read
\begin{align}
T_{\text{Sch.}}&=\frac{\hbar c}{4\pi k_B\mathcal{R}}\label{5.3}\\
T_{\text{QBH}}&=\frac{\hbar}{4\pi k_B\left(\frac{\mathcal{R}}{c}+\frac{\hbar\tilde{\omega}}{2Mc^2}\right)}\simeq\frac{\hbar c}{4\pi k_B\mathcal{R}}\left(1-\frac{\hbar\tilde{\omega}}{2M\mathcal{R}c}\right)~.\label{5.4}
\end{align}
We need to express $P_{\text{exc}}(\nu)d\nu$ as $P_{\text{exc}}(\lambda)d\lambda$ to truly investigate the Wien's law.  Throughout our analysis, we have considered $\nu\ll\Omega$ which implies  $\Omega\lambda\gg1$. Keeping in mind the above approximation, we can write down the $P(\lambda)d\lambda$ factor  from eq.(s)(\ref{5.1},\ref{5.2}) as
\begin{align}
P_{\text{exc}}^{\text{Sch.}}(\lambda)d\lambda\simeq&\frac{4\pi\mathcal{G}^2\mathcal{R}}{c\lambda^3\Omega^2}\frac{d\lambda}{e^{\frac{\hbar}{\lambda k_BT_{\text{Sch.}}}}-1}\label{5.5}\\
P_{\text{exc}}^{\text{QBH}}(\lambda)d\lambda\simeq&\frac{4\pi\mathcal{G}^2}{\lambda^3\Omega^2}\left(\frac{\mathcal{R}}{c}+\frac{\hbar\tilde{\omega}}{2Mc^2}\right)\frac{d\lambda}{e^{\frac{\hbar}{\lambda  k_BT_{\text{QBH}}}}-1}~.\label{5.6}
\end{align}
It is very straightforward to see that both the terms have similar dependence on the photon wavelength which is given by the following two functions
\begin{align}
D_{\text{Sch.}}(\lambda)&=\lambda^3\Omega^2\left(e^{\frac{\hbar}{\lambda k_B T_{\text{Sch.}}}}-1\right)\label{5.7}\\
D_{\text{QBH}}(\lambda)&=\lambda^3\Omega^2\left(
e^{\frac{\hbar}{\lambda k_BT_{\text{QBH}}}}-1\right)~.\label{5.8}
\end{align}
We want to find out the value of the wavelength of the scalar photon for which the excitation probability becomes maximum. For doing so, we start by finding out the $\lambda$ value for which eq.(s)(\ref{5.7},\ref{5.8}) becomes minimum. Calculating $\frac{d D(\lambda)}{d\lambda}$ and setting it equal to zero we, obtain the following two relations
\begin{align}
\frac{d D_{\text{Sch.}}(\lambda)}{d\lambda}=0\implies 1-e^{-\frac{\hbar}{\lambda k_B T_{\text{Sch.}}}}=\frac{\hbar}{3\lambda k_B T_{\text{Sch.}}}\label{5.9}\\
\frac{d D_{\text{QBH}}(\lambda)}{d\lambda}=0\implies 1-e^{-\frac{\hbar}{\lambda k_B T_{\text{QBH}}}}=\frac{\hbar}{3\lambda k_BT_{\text{QBH}}}\label{5.10}~.
\end{align}
Numerically solving eq.(s)(\ref{5.9},\ref{5.10}), we arrive at the following two results
\begin{equation}\label{5.11}
\lambda_{\text{Sch.}} T_{\text{Sch.}}\simeq \frac{0.3544 \hbar}{k_B},~\lambda_{\text{QBH}} T_{\text{QBH}}\simeq\frac{0.3544 \hbar}{k_B}~.
\end{equation}
It is important to note that for the above $\lambda$ values, $D_{\text{Sch.}}(\lambda)$ as well as $D_{\text{QBH}}(\lambda)$ becomes minimum and as a result both of the transition probabilities hit a maximum value. At the same time eq.(\ref{5.11}) indicates $\lambda T=\text{constant}$, confirming Wien's displacement law. Now considering eq.(\ref{3.20}) and neglecting the $\frac{\nu}{\Omega}$ terms in the coefficient factor, we arrive at the same displacement law given as
\begin{equation}\label{5.12}
\lambda_{\text{QCBH}}T_{\text{QCBH}}=\frac{0.3544 \hbar}{k_B}
\end{equation}
where the thermal bath temperature corresponding to the charged-quantum corrected geometry reads
\begin{equation}\label{5.13}
\begin{split}
T_{\text{QCBH}}&=\frac{\hbar}{4\pi k_B\left(\frac{\mathcal{R}}{c}\frac{(1+q)^2}{4q}+\frac{\hbar\tilde{\omega}(1+q^2)}{4Mc^2q^3}\right)}\\&\simeq \frac{\hbar q c}{\pi k_B\mathcal{R}(1+q)^2}\left(1-\frac{\hbar\tilde{\omega}}{\mathcal{R}Mc q^2}\frac{1+q^2}{(1+q)^2}\right)~.
\end{split}
\end{equation}
We shall now compare the three wavelengths using eq.(s)(\ref{5.11},\ref{5.12}) and plot it against the mass of the black hole. It is important to note that $\frac{\hbar\tilde{\omega}}{\mathcal{R}Mc}$ as a hole is a dimensionless constant factor and as a result we can set $\tilde{\omega}$ in such a way that the entire contribution is less than unity. We have set $\tilde{\omega}$ to be equal to $10^{33}$ to amplify the effect of the quantum gravity factor. For plotting $\lambda_{\text{QCBH}}$ against the mass of the black hole, we set the charge $q$  of the black hole to $e=0.06$ C. In general, $\tilde{\omega}$ is a very small factor, and the enhancement in the value of this parameter is done solely due to the amplification of the quantum gravity effect. The plot of $\lambda$ vs the mass of the black hole is given in Fig.(\ref{Lambda}).
We observe from Fig.(\ref{Lambda}) that with increasing charge, the critical wavelength value deviates more for lower-mass black holes and approaches the zero charge case for black holes with higher masses. The $q$ parameter restricts the charge and mass of the black hole to vary independently. For a black hole with a fixed mass $M$, the charge must obey the following relation $e\leq \sqrt{4\pi\varepsilon_0 G}M$ which also puts a restriction on arbitrarily using the charge value while plotting Fig.(\ref{Lambda}).
\linebreak
\section{Conclusion}\label{S6}
\noindent In this paper we consider the simple gedanken experiment where a two-level atom is freely falling into the event horizon of a quantum-corrected and charged black hole. Previously in \cite{OTM}, we analyzed this acceleration radiation due to freely falling atoms in the quantum-corrected Schwarzschild black hole geometry. It is important to note that the black hole is covered by a mirror in order to prevent the emitted radiation from interacting with the Hawking radiation from the black hole. Now the freely falling atoms emit acceleration radiation and we calculate the corresponding transition probability. Here we follow a slightly different approach in order to obtain the atomic trajectories where the radial coordinates in the integrand have been expressed in terms of the distance from the event horizon of the black hole together with the outer event horizon radius which is a constant. We finally obtain the transition probability of the atom for going from its ground state to a higher excited state and emitting a virtual photon simultaneously. We observe that the exponential term in the Planck factor of the transition probabilities picks up corrections proportional to $\frac{e^2}{M}$ which is absent in the general Reissner-Nordstr\"{o}m black hole without any quantum gravity correction. In the absence of this quantum gravity correction, the charge contribution comes in the fourth order of electrical charge and all quadratic order charge contributions cancel away. This is one of the most important observations in our work. We have then moved towards obtaining the horizon-brightened acceleration radiation entropy or the HBAR entropy for a quantum-corrected charged black hole. We have considered a cloud of two-level atoms falling freely into the event horizon of the quantum-improved charged black hole. We have then calculated the HBAR entropy by the quantum statistical mechanics approach and found that the entropy contains the Bekenstein-Hawking entropy term with subleading correction proportional to the logarithmic in the area of the black hole. Exactly similar subleading corrections have also been observed earlier in \cite{OTM} where we have claimed that logarithmic contributions in the HBAR entropy can only originate from such quantum-gravity corrections. Our current analysis reinforces the claim made in \cite{OTM} and proves that for quantum-improved black hole geometries, logarithmic corrections are universal.  Finally, we investigate Wien's displacement law in the case of the quantum-corrected charged black hole and compare it with the Schwarzschild as well as the quantum-corrected black hole case. We have then plotted the critical value of the wavelength (for which the excitation probability becomes maximum) against the mass of the black hole and observed higher deflections in the lower mass region for black holes with higher charge values.

\end{document}